\documentclass[12pt,twoside]{article}   
\usepackage[super,sort,comma]{natbib}

\usepackage{fancyhdr}		

\newcommand{\captionv}[3]{\begin{center}\parbox{#1cm}{\caption[#2]{{\sf #3}}}
\end{center}}

\usepackage[section]{placeins}   %

\usepackage{graphicx}

\makeatletter \renewcommand\@biblabel[1]{$^{#1}$} \makeatother
\newcommand{\cen}[1]{\begin{center} #1 \end{center}}

\usepackage[mathlines]{lineno}

\usepackage{hyperref}
\hypersetup{ colorlinks,
    citecolor=blue,
    filecolor=blue,
    linkcolor=blue,
    urlcolor=blue
}

\usepackage{xcolor}

\definecolor{gray}{rgb}{0.6,0.6,0.6}
\definecolor{red}{rgb}{0.85,0,0}
\definecolor{green}{rgb}{0,0.85,0}
\definecolor{blue}{rgb}{0,0,0.85}
\definecolor{beige}{rgb}{0.92,0.87,0.78}
\usepackage[all]{hypcap}    

\begin{document}

\cen{\sf {\Large {\bfseries Experimental validation of Fastcat kV and MV cone beam CT (CBCT) simulator} \\  
\vspace*{10mm}
Jericho O'Connell, Clayton Lindsay, and Magdalena Bazalova-Carter} \\
University of Victoria,  3800 Finnerty Road, Canada
\vspace{5mm}\\
Version typeset \today\\
}

\pagenumbering{roman}
\setcounter{page}{1}
\pagestyle{plain}
Author to whom correspondence should be addressed. email: jerichoo@uvic.ca\\

\begin{abstract}
\noindent {\bf Purpose:} To experimentally validate the Fastcat cone beam CT (CBCT) simulator against kV and MV CBCT images acquired with a Varian Truebeam linac.\\
{\bf Methods:} kV and MV CBCT images of a Catphan 504 phantom were acquired using a 100 kVp beam with the on-board imager (OBI) and a 6 MV treatment beam with the electronic portal imaging device (EPID), respectively. The kV Fastcat simulation was performed using detailed models of the x-ray source, bowtie filter, a high resolution voxelized virtual Catphan phantom, anti-scatter grid, and the CsI scintillating detector. Likewise, an MV Fastcat CBCT was simulated with detailed models for the beam energy spectrum, flattening filter, a high resolution voxelized virtual Catphan phantom, and the GOS scintillating detector. Experimental and simulated CBCT images of the phantom were compared with respect to HU values, contrast to noise ratio (CNR), and dose linearity. Detector modulation transfer function (MTF) for the two detectors were also experimentally validated. Fastcat’s dose calculations were compared to MC dose calculations performed with Topas. \\
{\bf Results:} For the kV and MV simulations, respectively: Contrast agreed within 14 and 9 HUs and detector MTF agreed within 4.2\% and 2.5\%. Likewise, CNR had a root mean squared error (RMSE) of 2.6\% and 1.4\%. Dose agreed within 2.4\% and 1.6\% of MC values. The kV and MV CBCT images took 71 and 72 seconds to simulate in Fastcat with 887 and 493 projections, respectively.  \\
{\bf Conclusions:} We present a multi energy experimental validation of a fast and accurate CBCT simulator against a commercial linac. The simulator is open source and all models found in this work can be downloaded from \href{https://github.com/jerichooconnell/fastCATs.git }{https://github.com/jerichooconnell/fastCAT.git}.\\


\end{abstract}

\newpage     

\tableofcontents

\newpage

\setlength{\baselineskip}{0.7cm}      

\pagenumbering{arabic}
\setcounter{page}{1}
\pagestyle{fancy}
\section{Introduction}

Cone beam computed tomography (CBCT) \cite{Jaffray2000Cone-beamCharacterization,Jaffray2002Flat-panelTherapy} is used extensively to visualize patient anatomy in the planning and administration of radiotherapy treatment. As such, CBCT that provides more accurate and realistic images can improve radiotherapy patient positioning and beam alignment. To this end CBCT is being constantly ameliorated, active research is being conducted end to end in the imaging process. To evaluate the efficacy of new research, two approaches are often used. New CBCT imaging methods can be evaluated based on an experimental phantom of known composition and dimensions with reproducible results but limited relation to real patient data. Conversely, evaluations can be performed using real patient data of unknown composition and dimensions with limited reproducibility. The ideal case, to evaluate new research on a realistic patient model with known composition, dimensions and robust reproducibility is an active area of study and the focus of this work.

With the introduction of new anthropomorphic virtual phantoms \cite{Xu2011DynamicStudy,Lee2005ThePatients,Zhang2006ChineseProject,Kramer2010FASHCalculations,Segars20104DResearch}, a new approach to evaluating x-ray imaging research has become viable. Realistic anthropomorphic phantoms allow simulations to be performed with realistic patient anatomy without the patient dose and uncertain ground truth associated with clinical data. Numerous simulators exist that are compatible with these phantoms.

These CT simulators generally use Monte Carlo (MC) platforms \cite{Badal2009AcceleratingUnit,Li2011Patient-specificPatients,Jiang2004AdaptationData}. These MC platforms, while used primarily for dose calculation in radiation therapy, have superior accuracy due to precise simulation of the particle transport underlying CBCT. However, MC methods suffer from long compute times for simulating CBCT. This is due to a variety of factors, e.g. the number of projections needed, the handling of a high resolution voxelized phantom, and the transport of optical photons in the simulator, all of which can be detrimental to computation speed. For example, in a study by Blake \textit{et al.} \cite{Blake2013CharacterizationGeant4} simulation of one projection image with a scintillating detector and 10$^7$ primary x-rays took 3000 CPU-hours. Likewise, MC methods suffer from increased load times, compute times, and memory constraints as the resolution of an anthropomorphic phantom is increased. For example, in the work of Yeom \textit{et al.} \cite{Yeom2019ComputationPHITS}, which examines the performance of Geant4, MCNP, and PHITS with respect to dose calculation speeds in voxelized anthropomorphic phantoms. Simulation times were increased by an order of magnitude in Geant4 when decreasing voxel size from 1 mm$^3$ to 0.1 mm$^3$, while MCNP nor PHITS could simulate a 0.1 mm$^3$ phantom due to the number of phantom voxels exceeding the maximum limit of the simulation codes.

Other simulation approaches are currently available to improve simulation time. Graphical processing unit (GPU) MC is a current method that shows speed up factors of 27-90 over traditional CPU-based MC \cite{Bert2013Geant4-basedApplications,Badal2009AcceleratingUnit}. Likewise, ray-tracing methods are available for CPU and GPU \cite{vanderHeyden2018VOXSI:Imaging,Landry2013ImaSimRadiology}. These methods are very fast but cannot account for scatter in phantoms and detector response. A variety of hybrid methods also exist \cite{Jia2012AProjections,DeMan2007CatSim:Environment,Li2008AImaging,Abadi2019DukeSim:Tomography}. Hybrid methods have been successful in some applications. For example, gDDR simulations \cite{Jia2012AProjections} showed agreement with experimental noise amplitude in CBCT with a relative difference 3.8\%. However, these methods have some limitations. The methods were either not validated against commercial CBCT scanners for realistic voxelized phantoms or take hours to generate a CBCT image without a high performance computing cluster.

Here we experimentally validate Fastcat, an efficient hybrid CBCT simulator of both kV and MV CBCT imaging, previously validated by means of MC simulations \cite{OConnell2020FastCAT:Simulation}. This simulation platform is capable of simulating both kV and MV commercial CBCT scanners and takes a few minutes to run a CBCT simulation with the same number of projections as a commercial scanner on a desktop computer with a single GPU. A comprehensive evaluation of image quality metrics is performed to demonstrate the accuracy of the method. This platform enables rapid evaluation of new CBCT protocols. 

\section{Materials and Methods}

\subsection{Experimental Data Acquisition}

\begin{figure}[h!]
  \begin{center}
  \includegraphics[width=0.75\textwidth,trim={1cm 4cm 9.5cm 2cm}, clip]{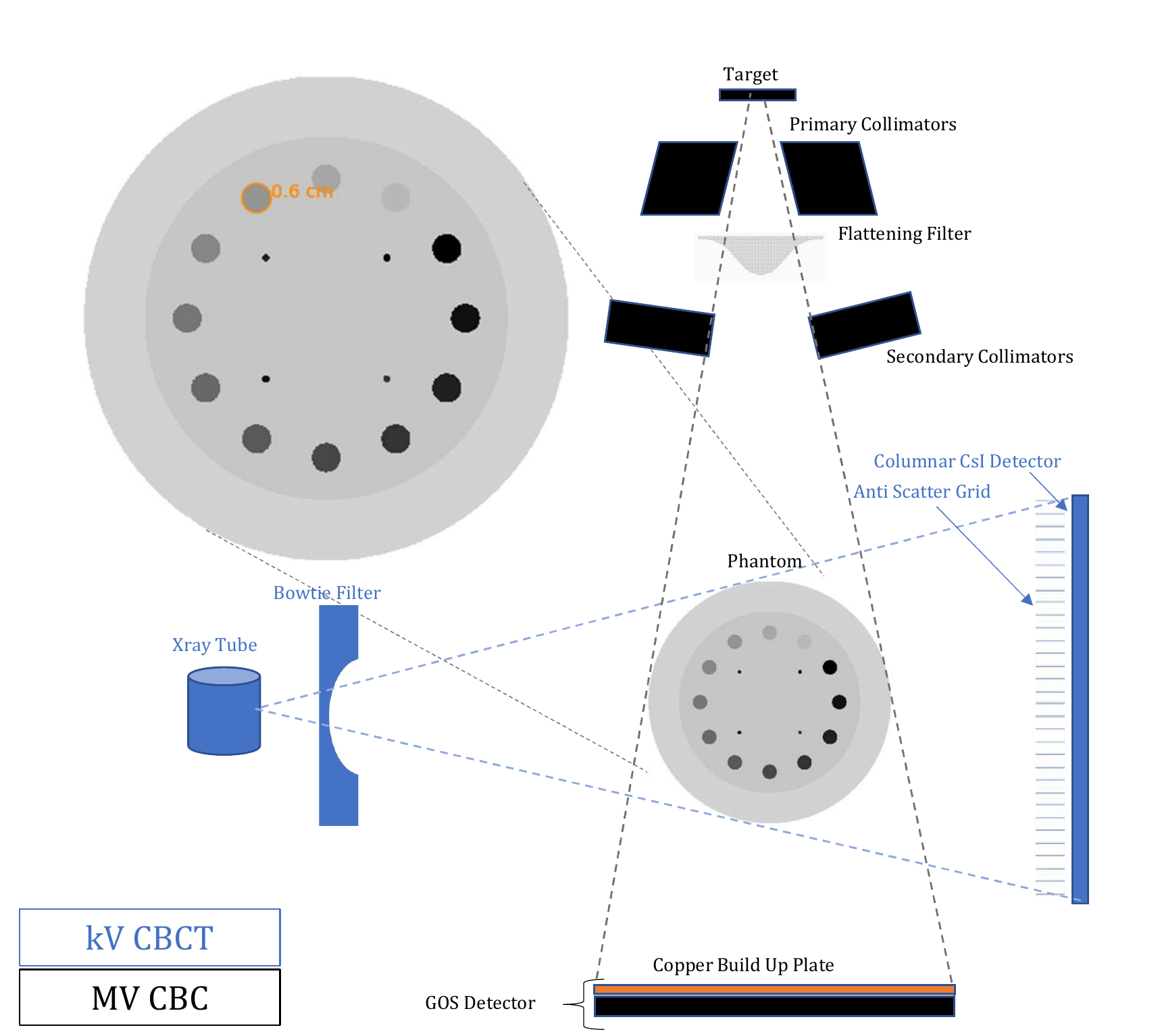}
  \captionv{12}{}
  {A schematic of the Truebeam linac and CTP404 module. The kV and MV imaging systems are shown in blue and black, respectively. Dotted lines outline where the beams would be generated. The phantom displayed is used to represent the phantom geometry. Thus, the phantom colormap does not correspond to any material property. The schematic is not to scale. 
  \label{fig_setup} 
    }  
    \end{center}
\end{figure}

Fastcat was validated for two different experimental CBCT data sets acquired on a Truebeam STx linac (Varian Medical Systems, Palo Alto, CA).

\subsubsection{kV CBCT}

An experimental kV CBCT image was acquired using the Varian Truebeam kV Planar Imager (which we also refer to as the OBI). The kV CBCT image was acquired using a Varian GS-1542 x-ray tube with a 100 kVp x-ray beam. The spectra was filtered by 2.7 mm aluminum inherent filtration as well as a 0.89 mm titanium beam hardening filter. An aluminum bowtie filter was also used with a minimum thickness of 1.53 mm and a maximum thickness of 27.42 mm. The detector was a PaxScan 4030CB (Varex Imaging Corporation) flat panel detector with an anti-scatter grid (ASG). The ASG was the default model with a grid ratio of 10, septal thickness of 0.036 mm, and line density of 60 lines/cm. The detector itself was a 0.6 mm columnar CsI detector with a fill factor of 70\% and an amorphous silicon readout pitch of 194 $\mu$m. The scan was acquired in the flouro-mode with pixels binned to 388 $\mu$m. Thus, the detector had a pixel matrix of 1024$\times$768, with a total detector size of 397$\times$298 mm. A total of 887 views of the phantom were acquired at equally spaced angles over a 360 degree rotation. The scan CTDI$_{vol}$ was 21.1 mGy. The CBCT was acquired in developer mode and the reconstruction was performed using the FDK algorithm \cite{Feldkamp1984PracticalAlgorithm} with a ram-lak filter. The reconstruction kV CBCT image size was 512$\times$512 pixels with isotropic voxel dimensions of 0.391 mm$^3$.

\subsubsection{MV CBCT}

The MV CBCT was acquired using the 6 MV therapy beam with the default tungsten target and a flattening filter. The detector used was a Varian as1200 GOS detector with a pixel pitch of 0.392 mm. Optical properties of the detector are described in more detail in the work of Shi \textit{et al.} \cite{Shi2018APerformance}. The detector had a pixel matrix of 1024$\times$768, with a total detector size of 401$\times$301 mm. A total of 493 views of the phantom were taken at equally spaced angles over a 360 degree rotation. The scan was acquired with a dose of 300 MU. A high dose was used to ensure adequate image quality from the low quantum efficiency GOS detector. The CBCT was acquired in developer mode and the reconstruction was performed using the FDK algorithm with a ram-lak filter. The reconstructed CBCT image size was 512 $\times$ 512 pixels with isotropic voxel dimensions of 0.391 mm$^3$. An fft-wavelet ring artifact reduction was used to reduce artifacts in the presented image \cite{Munch2009StripeFiltering}. Ring artifact reduction was not used in the analysis since the dampening of high wavelet frequencies can reduce image noise.

\subsection{Fastcat Overview}

The goal of Fastcat is to simulate CBCT with the highest computational efficiency without compromising accuracy. Fastcat’s methodology is to separate a MC simulation into primary radiation and scatter. The primary radiation is attenuated analytically while the scatter is simulated in MC. The scatter is then curve fit to remove noise. Scatter is only calculated once per phantom and stored in Fastcat. To allow for arbitrary polyenergetic beams the primary and secondary radiation are both simulated at 18 discrete energies between 10 keV to 6 MeV. To form the image the primary radiation and scatter are weighted and combined. These weights are determined by the input spectrum and the detector energy response. A detailed description of Fastcat can be found in the previous work of O’Connell and Bazalova-Carter \cite{OConnell2020FastCAT:Simulation}. Here we will focus on the descriptions of additional Fastcat features that were included in this work.

Fastcat simulations were performed on a linux desktop computer with 16 GB memory a Nvidia GeForce RTX 2070 GPU and eight Intel Skylake CPUs.

\subsubsection{Phantom}
The phantom used in this study was the CTP404 sensitometry module from a Catphan 504 phantom (The Phantom Laboratories, Salem, NY). The 20-cm diameter sensitometry module contains 1.2-cm diameter inserts filled with air, acrylic, low-density polyethylene (LDPE), teflon, polystyrene, delrin and polymethylpentene (PMP). The phantom was positioned with the center of the CTP404 module at isocenter.

\subsection{Fastcat CBCT Simulations}


The kV Fastcat simulation used as input an analytical x-ray source generated in xpecgen \cite{Hernandez2016Xpecgen:Anodes} with an anode angle of 14$^{\circ}$ and a tube voltage of 100 kVp. The source was subject to additional analytical filtration of 2.7 mm of aluminum and a 0.89 mm titanium beam hardening filter. The detector modelled in the simulation was a columnar CsI detector with optical properties based on the work of Freed \textit{et al.} \cite{Freed2009ExperimentalScreens}. The columnar CsI was 0.6 mm thick with a fill factor of 70\% and a pixel pitch of 384 $\mu$m.


The MV Fastcat simulation source was a 6 MV Truebeam phase-space file provided by Varian. This phasespace was binned by photon energy to provide a photon spectra for input into Fastcat. The detector used in the simulation was modelled as a Varian as1200 detector using the optical properties from Shi et al \cite{Shi2018APerformance}. This detector is the same as described in previous work \cite{OConnell2020FastCAT:Simulation}. The GOS scintillator was 0.29 mm thick with a pixel pitch of 392 $\mu$m.


\subsection{Additional Fastcat Modelling}

\subsubsection{Bowtie and Flattening Filters}

Additional models in Fastcat are applied to model some features of the scans. The kV bowtie filter was known to have a minimum thickness of 1.53 mm aluminum and a maximum thickness of 27.3 mm aluminum. The proprietary bowtie filter shape was unknown and estimated from the shape of a kV OBI air scan. With Fastcat’s model of the kV imager spectra and the energy response of the detector, a weighted attenuation coefficient of aluminum was calculated for the kV OBI. This attenuation coefficient was then used to estimate the thickness of aluminum filtration present at each pixel. While an additional thickness of 1.53 mm of aluminum was added to each pixel to account for the base thickness of the bowtie filter. The maximum thickness of aluminum was calculated to be 28.3 mm using this method. This thickness is seen to be nearly equivalent to the specified thickness plus an additional factor accounting for a non-perpendicular path through the filter. Fastcat’s kV beam was then filtered by the calculated amount of aluminum at each pixel during simulation. Additionally, the MC scatter for the kV image was recalculated to include the bowtie filter’s effect on scatter. To account for the heel effect in the beam, tungsten filtration in the shape of a linear ramp increasing in the cathode-anode direction with a maximum height of 0.1 mm was added.

The thickness of the MV flattening filter was also estimated. In this case, the Fastcat beam was attenuated by a gaussian tungsten flattening filter. The height and standard deviation were chosen such that a profile through the reconstructed CBCT best matched experimental CBCTs.

\subsubsection{Anti-scatter Grid}

To correctly model the kV OBI a model for the anti-scatter grid (ASG) was introduced. The model is based on measurements made by Wiegert \textit{et al.} \cite{Wiegert2004PerformanceCT} Primary transmission factor and scatter transmission factors from the 44r10 ASG discussed in the paper were used. These factors are 0.76 and 0.37, respectively. Primary fluence reaching the detector in the Fastcat simulation was multiplied by the primary transmission factor. The scatter transmission factor was used to separate the scatter into two portions handled by the simulation as either accepted or attenuated, respectively. The first portion was accepted by the detector while the second portion was assumed to be incident on the ASG at a 12 degree angle. Where 12 degrees was the mean angle of scatter incidence from Weigert \textit{et al.}’s work. This second portion was filtered by 173 $\mu$m of lead, which is the path length through the 36 $\mu$m lamella at an angle of 12$^\circ$. The number of particles in these two portions were calculated such that the total Fastcat scatter transmission factor agreed with the theoretical.

\subsubsection{Virtual Phantom}

 A virtual version of the Catphan phantom was created. The virtual phantom was voxelized into 1024 pixels horizontally (h) 1024 pixels vertically (v) with voxel dimensions of 0.195 mm (h), (v) and a slice thickness of 0.313 mm. Materials linear attenuation coefficients were calculated using the material compositions and densities in the Catphan documentation. Densities of Delrin and the inner and outer body materials of the phantom were estimated as data was not available on their exact composition. Delrin’s composition was estimated to match the attenuation coefficient available on the manufacturer site. The two body materials were estimated using the composition of acrylic and densities that would give the materials the relative electron densities reported in Star-Lack \textit{et al.} \cite{Star-Lack2015AImaging}.

\subsection{Dose Comparison}

A validation of dose linearity was performed. Four CBCTs at increasing doses were acquired experimentally for both the kV and the MV setups. Doses were measured using machine CTDI for the kV CBCTs with CTDIs of 5.27, 7.03, 10.55, and 21.1. Four MV CBCTs were acquired with machine MU values which were 75, 100, 150, and 300 MU. The dose in Fastcat was adjusted such that the average CNR over all inserts in the Catphan phantom agreed with that of the experimental CBCT with the lowest dose averaged. This Fastcat dose was then multiplied by factors of 1.5, 2, and 4. Agreement between average CNR was then compared between these Fastcat simulations and the experimental CBCTs.

An MC validation of the mean phantom dose was performed. Mean phantom dose was calculated in Topas \cite{Perl2012TOPAS:Applications} using an MC model of the Catphan phantom with 2$\times$10$^8$ initial photons. This dose was compared to the dose calculated in Fastcat for the same number of photons. These simulations used the Geant4 Penelope physics list and a particle range cutoff of 5$\mu$m. Simulations were run on a linux desktop computer with 8 Intel Skylake CPUs. No variance reduction techniques were used. Two simulations were run in total, one with the 6 MV Varian phasespace file and one with an analytical 100 kVp x-ray spectrum filtered by 2.7 mm of aluminum, 0.89 mm titanium, and an aluminum bowtie filter. Both simulation measured the mean phantom dose to virtual Catphan 504 phantom.

\subsection{Validation Metrics}

\subsubsection{Detector MTF}

Detector modulation transfer function (MTF) was measured using a slanted virtual slit of 0.3 mm in Fastcat. The method of measuring the detector MTF is discussed in previous work \cite{OConnell2020FastCAT:Simulation}. Detector MTF was compared to experimental results from previous works of Howansky \textit{et al.} and Shi \textit{et al.} \cite{Howansky2017DirectImaging,Shi2018APerformance}.

\subsubsection{CBCT Contrast and CNR}

To validate Fastcat, image quality was compared between Fastcat and experimental images. Both images were first converted to Hounsfield Units (HU) by subtracting the CT value of water in the image and scaling by the difference between water and air regions in the image. 

\begin{equation}
HU = 1000 \frac{im - \mu_{water}}{(\mu_{water} - \mu_{air})}
\end{equation}
                   
Where $im$ is the reconstructed image and $\mu_{water}$ and $\mu_{air}$ are the linear attenuation coefficients of water and air, respectively. For Fastcat, water values were taken from a separate scan of a water phantom. Contrast was compared using regions of interest (ROIs) in each insert of the CTP404 sensiometry module. Regions of interest were made slightly smaller than the inserts to avoid partial volume effects. The mean value over sixteen slices was used as the contrast value while the standard deviation of the contrast in the sixteen slices was used as an estimate of the standard deviation. Contrast to noise ratio (CNR) was measured using the same ROIs. The standard deviation of each ROI was used as an estimate of the noise. Contrast, again, was measured against water, the CNR was then

\begin{equation}
CNR = \frac{\mu_{ROI} - \mu_{water}}{(\sigma_{ROI}^2 + \sigma_{water}^2)^{1/2}}
\end{equation}

Where $\mu_{ROI}$ and $\sigma_{ROI}$ are the mean and standard deviation of each ROI. CNR was measured in each slice over sixteen slices, with the mean value used as a final estimate of the CNR. The standard deviation of the sixteen slices was used as an estimate of the standard deviation of the CNR.

\section{Results}

\subsection{CBCT Image Comparison}

\begin{figure}[ht!]
  \begin{center}
  \includegraphics[width=\textwidth, clip]{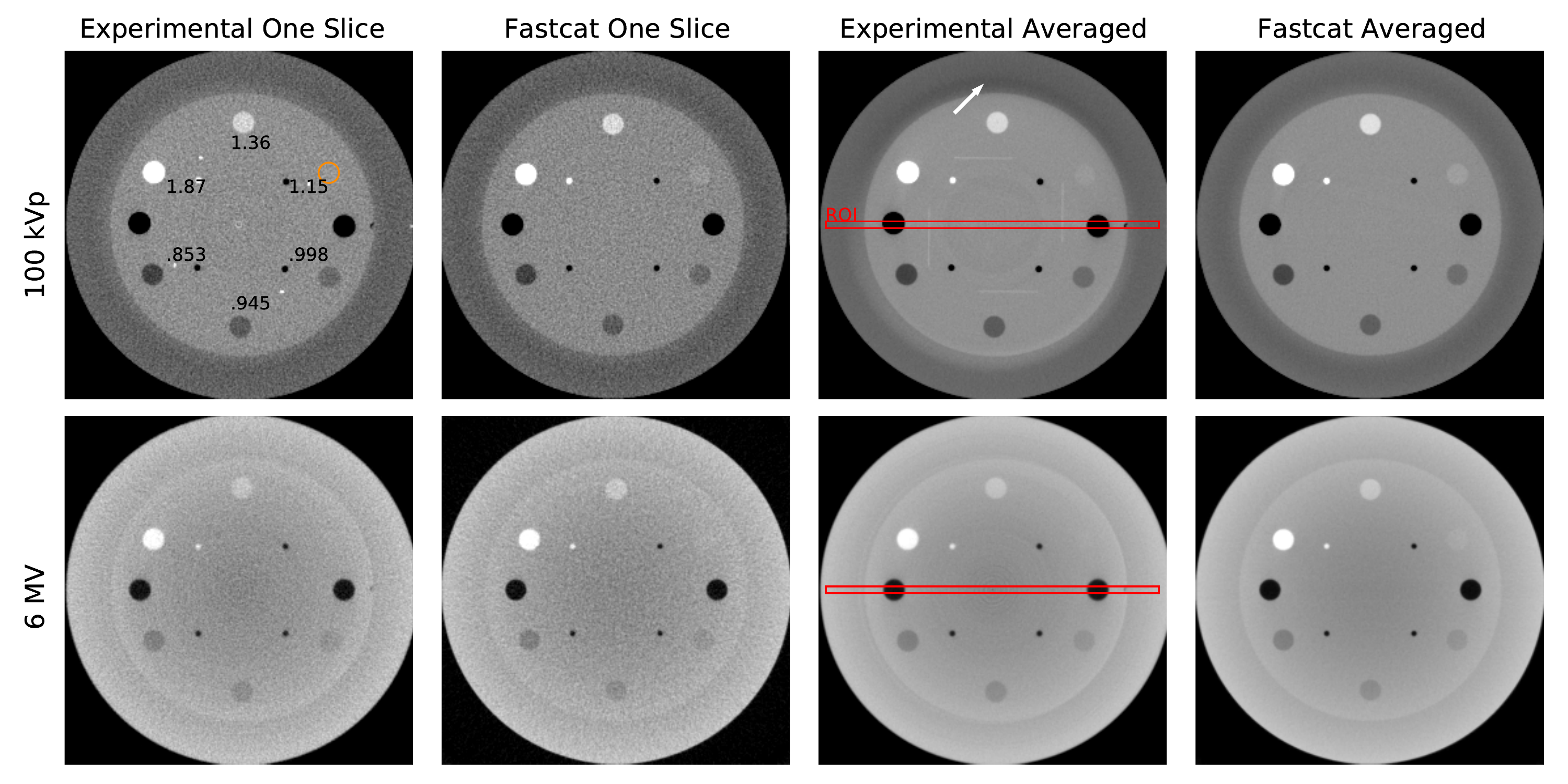}
  \captionv{12}{}
  {A qualitative comparison between experimental CBCTs and Fastcat simulations averaged over 16 slices and for one CBCT slice, respectively. The relative electron densities of the inserts are shown in black. The mean of the ROI shown in red are plotted in Figure \ref{profile}. An arrow indicates a crescent artifact from mechanical misalignment. Window (W) and level (L) of 1000/300. 
  \label{images} 
    }  
    \end{center}
\end{figure}

Experimental and Fastcat simulated CBCTs were compared. Figure \ref{images} shows a comparison of 100 kVp and 6 MV experimental and Fastcat images, respectively. Qualitatively, the experimental and Fastcat-simulated CBCT are similar. The effect of the bowtie filter can be seen in the 100 kVp CBCTs which are qualitatively uniform. The effect of the flattening filter can be seen in the 6 MV images where beam hardening artifacts can be seen.

\begin{figure}[ht!]
  \begin{center}
  \includegraphics[width=0.75\textwidth, clip]{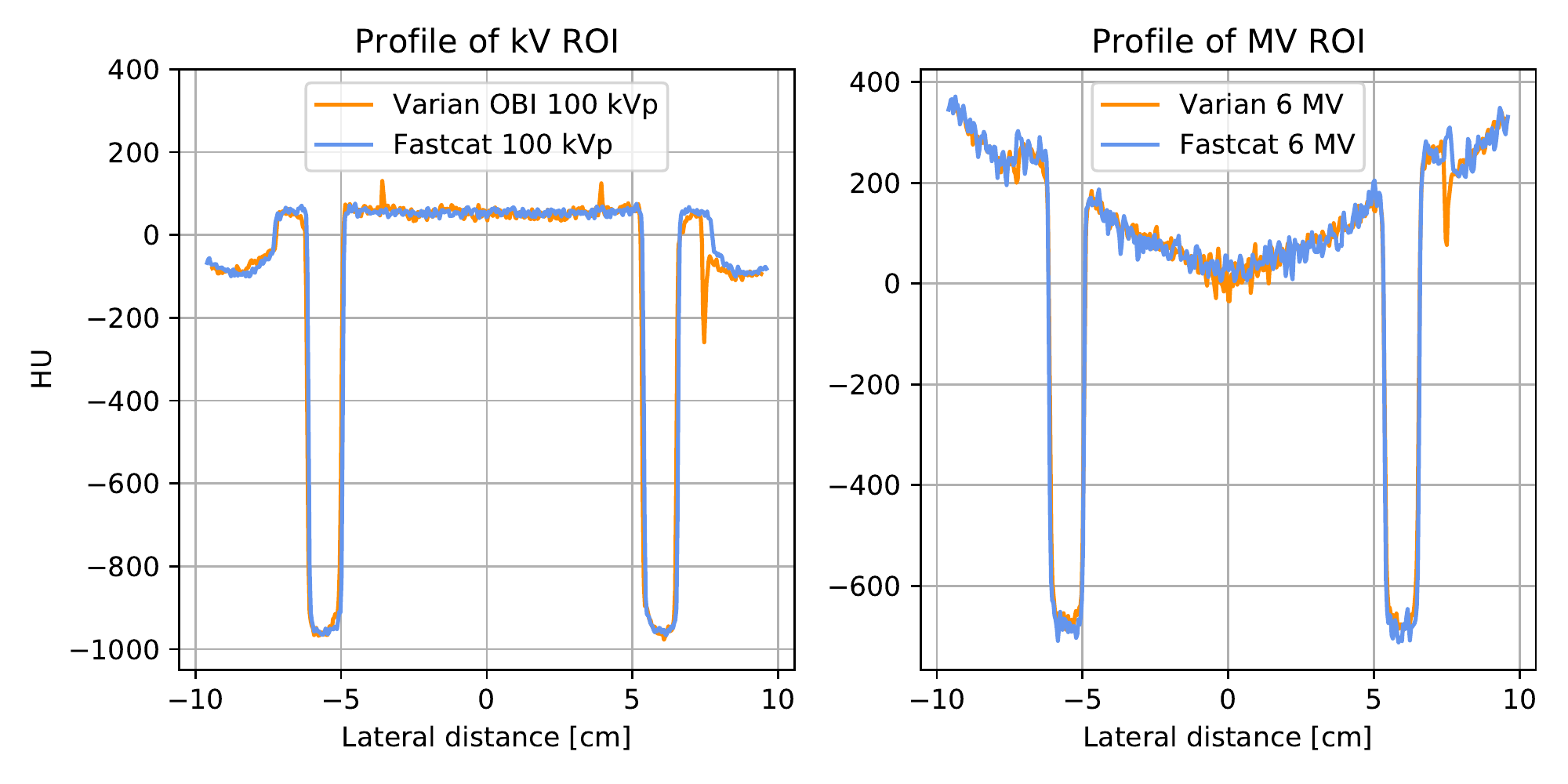}
  \captionv{12}{}
  {Profiles over the ROI marked in red in Figure \ref{images}.
  \label{profile} 
    }  
    \end{center}
\end{figure}

Different features can be seen in the averaged CT images in Figure \ref{images}. The 100 kVp image shows a radial uniformity thanks to the presence of the bowtie filter. However, the image is not completely uniform. The bowtie filter ends just inside the lower density outer layer of the Catphan. This causes non-uniformity in the phantom in this lower density region. A difference between the experimental and Fastcat kV images is a large crescent artifact denoted by a white arrow in Figure \ref{images}. The effect of this artifact can also be seen at the edges of the kV profile in Figure \ref{profile}. In the Varian documentation such artifacts are said to be related to mechanical misalignment between the detector, x-ray tube and bowtie filter. Since Fastcat does not account for this sort of artifact it is not present in the Fastcat image. 

In the 6 MV image, the flattening filter causes the beam to be harder in the center of the phantom than the periphery. This leads to cupping in the center of the phantom as seen in both images. Some difference is seen between MV profiles in the center of the phantom in Figure \ref{profile}. This difference may be from ring artifacts or due to the estimation of the flattening filter as Gaussian. 

Overall, the contrast is visibly similar in between both sets of images in Figure \ref{images}. The 100 kVp image has high contrast, clearly differentiating all of the inserts while the 6 MV image has low contrast reflecting the lower number of photo-electric interactions at this energy.

\begin{figure}[ht!]
  \begin{center}
  \includegraphics[width=0.75\textwidth, clip]{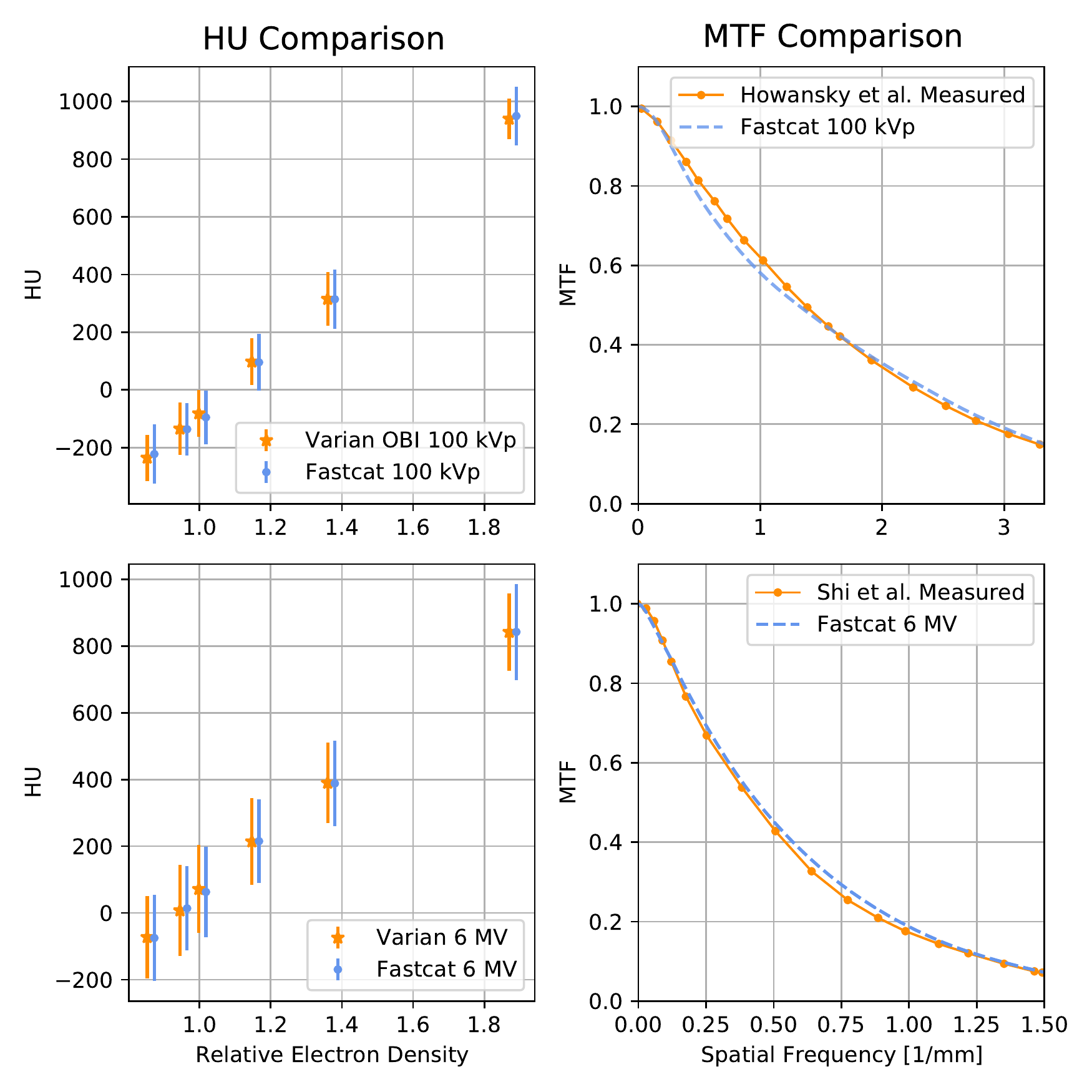}
  \captionv{12}{}
  {(L) A comparison is demonstrated between the experimental kV and MV CBCTs and the Fastcat simulations in terms of HU values in an experimental and virtual CTP404 module, respectively. (R) Comparison of detector MTFs calculated by Fastcat and based on experimental data presented by Howansky \textit{et al.} \cite{} and Shi \textit{et al.} \cite{Shi2018APerformance} for the kV and MV detector, respectively.
  \label{HU} 
    }  
    \end{center}
\end{figure}

\subsection{Contrast and MTF}

\begin{table}[]
\caption{kV and MV HU values in CTP404 module, values are in HU while standard deviations are in parentheses.
}
\label{table}
\begin{center}
\begin{tabular}{|l|c|c|c|c|c|c|}
\hline
                & PMP        & LDPE       & Polystyrene & Acrylic   & Delrin    & Teflon     \\ \hline
kV Exp.     & -204 (91)  & -102 (102) & -48 (92)    & 135 (91)  & 357 (105) & 995 (87)   \\ \hline
kV Fastcat  & -192 (103) & -104 (91)  & -62 (93)    & 133 (97)  & 356 (102) & 1005 (102) \\ \hline
Abs. Diff.  & 12         & 2          & 14          & 2         & 1         & 10         \\ \hline
MV Exp.     & -70 (124)  & 14 (136)   & 74 (131)    & 219 (129) & 397 (121) & 847 (115)  \\ \hline
MV Fastcat  & -73 (129)  & 17 (126)   & 65 (135)    & 222 (124) & 398 (128) & 850 (144)  \\ \hline
Abs. Diff.  & 3          & 3          & 1           & 3         & 1         & 3          \\ \hline
\end{tabular}
\end{center}
\vspace{10mm}
\end{table}

Looking quantitatively at these images, there is also close agreement. In Figure \ref{HU}, HU values are compared between experimental and Fastcat CBCT images. The HU values in the inserts are shown in Table \ref{table}. HU values for the kV and MV images agreed within 14 and 9 HU values, respectively. The kV HU values were all within the range defined in the Catphan documentation. kV and MV detector MTF was also compared to measurements made by Shi \textit{et al.} and Howansky \textit{et al.}, respectively. The measurements for the CsI detector were within 4.2\% of measurements by Howansky \textit{et al.} with an average RMSE of 1.7\%. MTF deviated mostly at frequencies between 0.5-1.3 mm$^{-1}$ while agreeing closely at larger spatial frequencies. The MTF measurements for the GOS detector were within 2.5\% of  measurements by Shi \textit{et al.} agreeing well over all spatial frequencies with an average root mean squared error (RMSE) of 0.5\% and a slightly higher MTF at spatial frequencies between 0.5-1.0 mm$^{-1}$.

\subsection{CNR and Dose}

\begin{figure}[ht!]
  \begin{center}
  \includegraphics[width=0.75\textwidth, clip]{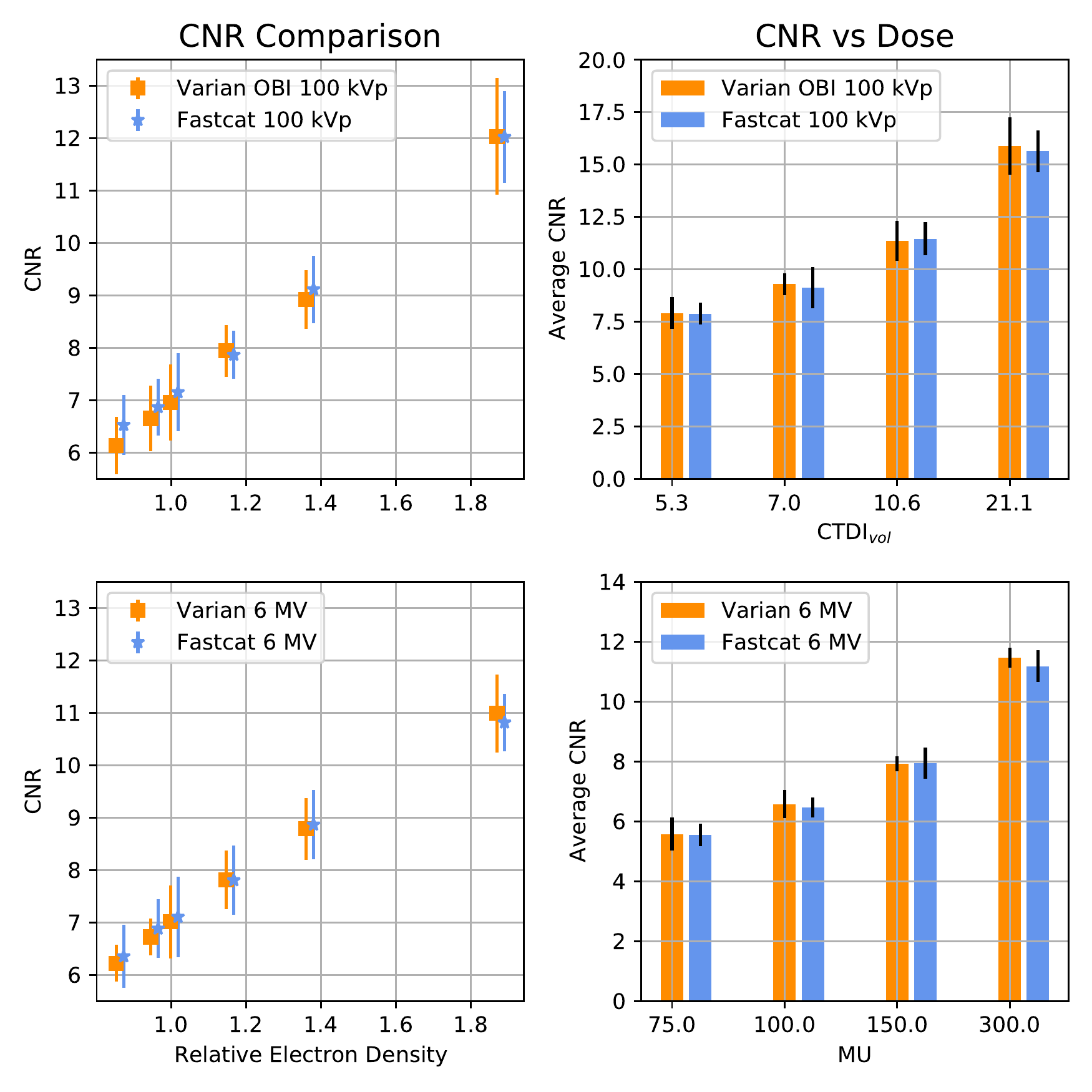}
  \captionv{12}{}
  {(L) A comparison is shown between the Varian Truebeam kV and MV CBCTs and the Fastcat simulated CBCTs in terms of CNR in an experimental and virtual CTP404 module, respectively. (R) CNR is compared between Truebeam CBCTs and Fastcat CBCTs as a function of dose. \label{CNR} 
    }  
    \end{center}
\end{figure}

CNR was examined in each insert for each of the images seen in Figure \ref{images}. The CNR results are shown in Figure \ref{CNR}. CNR values are summarized in Table \ref{tableCNR}. CNR agreed within 0.4 and 0.2 for the kV and MV images, respectively. CNR as a function of dose was examined as seen in Figure \ref{CNR}. Fastcat CNR values were seen to have average RMSE of 2.6\% and 1.4\% for kV and MV images, respectively. RMSE between experimental and Fastcat CNRs as a function of dose were seen to be 1.19\% and 1.22\%, respectively.

\begin{table}[h!]
\caption{kV and MV CNRs in CTP404 module, standard deviations are in parentheses.
}
\label{tableCNR}
\begin{center}
\begin{tabular}{|l|l|l|l|l|l|l|}
\hline
           & PMP       & LDPE      & Polystyrene & Acrylic   & Delrin    & Teflon      \\ \hline
kV Exp.    & 6.1 (.55) & 6.7 (.62) & 6.9 (.72)   & 7.9 (.50) & 8.9 (.56) & 12.0 (1.11) \\ \hline
kV Fastcat & 6.5 (.57) & 6.9 (.54) & 7.1 (.74)   & 7.9 (.46) & 9.1 (.64) & 12.0 (.88)  \\ \hline
Abs. Diff. & 0.4       & 0.2       & 0.2         & 0.0       & 0.2       & 0.0         \\ \hline
MV Exp.    & 6.2 (.34) & 6.7 (.35) & 7.0 (.70)   & 7.8 (.56) & 8.7 (.59) & 11.0 (.74)  \\ \hline
MV Fastcat & 6.4 (.60) & 6.9 (.56) & 7.1 (.76)   & 7.8 (.65) & 8.9 (.66) & 10.8 (.55)  \\ \hline
Abs. Diff. & 0.2       & 0.2       & 0.1         & 0.0       & 0.2       & 0.2         \\ \hline
\end{tabular}
\end{center}
\vspace{10mm}
\end{table}

\subsection{Dose and Speed}

Dose was compared between MC simulations and Fastcat. Dose was calculated for both the kV and the MV CBCT aquisitions. The mean doses to the entire Catphan for the kV beam were 0.416 and 0.426 $\mu$Gy for the MC and Fastcat calculation, respectively. The mean doses to the Catphan for the MV beam were 7.31 and 7.19 $\mu$Gy for the MC and Fastcat calculation, respectively. Thus, there were differences of 2.4\% and 1.6\% between Fastcat and MC dose calculations for kV and MV, respectively.

\begin{figure}[h!]
  \begin{center}
 
  \includegraphics[width=0.5\textwidth, clip]{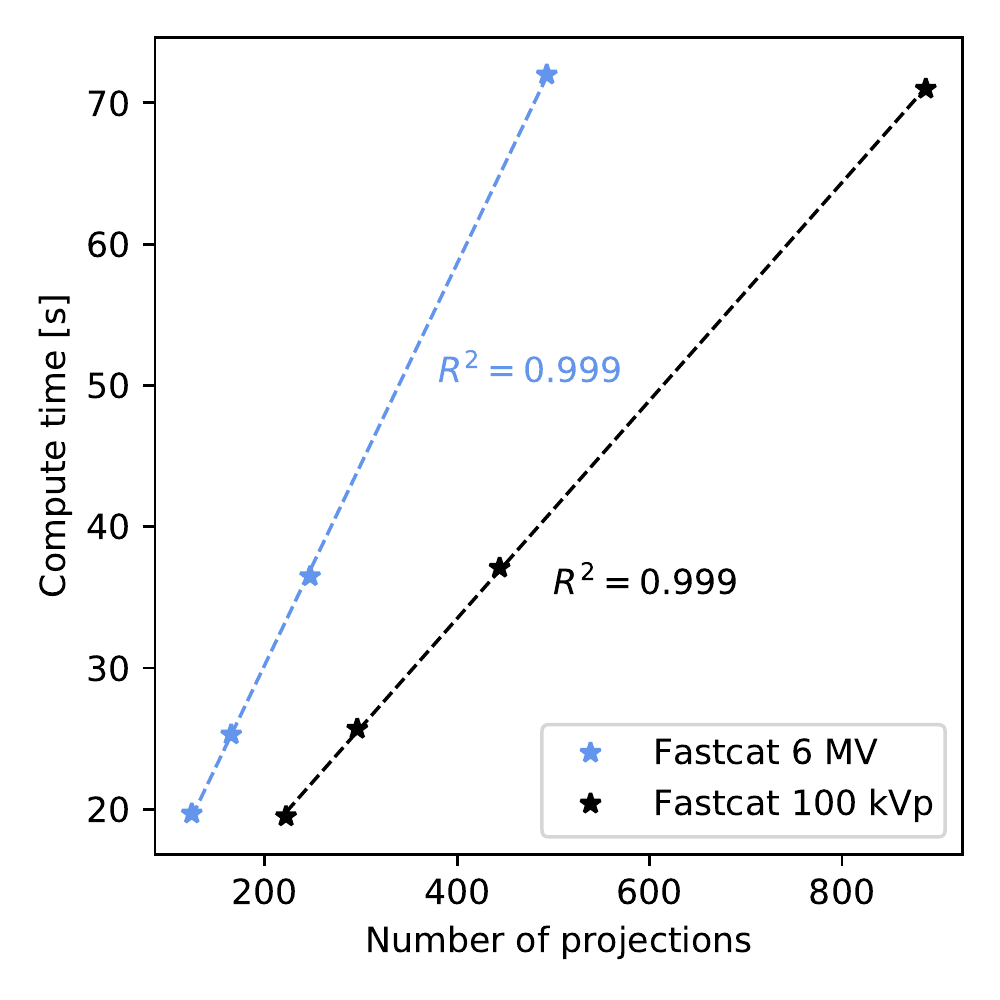}
  \captionv{12}{}
  {A time comparison as a function of the number of projections in a Fastcat CBCT simulation. 
 \label{speed} 
    }  
    \end{center}
\end{figure}

Simulation times were recorded for four Fastcat simulations to estimate the relationship between computation time and the number of projections. Results are shown in Figure \ref{speed}. Computation time scaled linearly with the number of projections for the MV and kV Fastcat simulations with both linear fits having R$^2$ values above 0.999. kV simulations are faster, scaling at 0.37 s/projection while MV simulations scaled at 0.55 s/projection. kV simulations are seen to be more computationally efficient since only ten discrete energies are simulated between 10 and 100 keV while the MV simulation used 18 energies between 10 keV and 6 MeV.

\section{Discussion}

We present and perform a two energy validation of Fastcat, a free open-source CBCT simulator against CBCTs acquired on a Varian Truebeam linac. This simulator performed simulations of high-resolution voxel based phantoms on a desktop computer with 16 GB of memory and an Nvidia RTX 2070 GPU. Simulation times were on the order of a minute to simulate the images used in the validation. Simulated CBCT images demonstrate unprecedented accuracy for their speed, enabling novel application of CBCT simulation in research and clinical workflows. A comparison of Fastcat to four other simulation tools is discussed. DukeSim, CatSim, VOXSI, and gDDR were selected for comparison since they accept voxelized phantoms, use analytical methods to reduce simulation times, and were validated with experimental data.

Compared to the validation of other CT and CBCT simulators, the experimental validation of Fastcat is exceptional in its comprehensiveness and accuracy. The two energy validation presented, using very different imaging systems mitigates possible overfitting resulting from a validation on a single scanner as was conducted for DukeSim, CatSim, and gDRR. VOXSI provides an experimental validation at three tube potentials, however, it uses the same source and detector in each simulation. This might lead to uncertainty in extrapolating the simulator to different detectors and sources, as simulation parameters may be unintentionally specific to the one setup.

Fastcat was experimentally validated with the ubiquitous Catphan 504 phantom and standard imaging metrics such as HU values and CNR. Of the simulation tools discussed, only VOXSI is experimentally validated using a common phantom while other tools were validated using only a single material. VOXSI simulations were within 42 HU of a Gammex RMI 467 tissue characterization phantom (Gammex, Middleton, WI), while Fastcat which was within 14 HU of a comparable Catphan CTP404 module. DukeSim displayed simulated CT images of a Mercury phantom \cite{Wilson2013ASystems}, however neither contrast nor CNR comparisons were shown. This Fastcat validation presents CNR agreement to experimental data with an average RMSE of 2.6\% and 1.4\% in the CTP404 module inserts for kV and MV images, respectively.

Unique among the simulation tools discussed, Fastcat demonstrates agreement with detector MTF. Values were within 4.2\% and 2.5\% of measurements for CsI and GOS detectors, respectively. CatSim and DukeSim use a simple estimation of detector spatial resolution based on neighbouring crosstalk. In reality, scintillating detectors have crosstalk that extends further than their nearest neighbors. Fastcat accounts for this based on the fastEPID framework \cite{Shi2019ADetectors.} which found an 81 $\times$ 81 pixel OSF was necessary to simulate an accurate detector response. VOXSI and gDRR do not include simunelations of the spatial resolution.

Computational efficiency was paramount in Fastcat’s methodology. As such, Fastcat’s computational efficiency compares favorably to other CT and CBCT simulation tools. Direct comparisons currently remain unfeasible as simulation times depend on phantom resolution, processors, noise, and the number of projections, variables which are not always presented in the literature. Nevertheless, qualitative comparisons can be made by examining reported simulation times: gDDR and Fastcat are estimated to have similar simulation speeds, 360 projections of a gDRR high resolution anthropomorphic head phantom  took 95.3 s on an Nvidia GTX 590 GPU. Analogously, Fastcat took 28 seconds to simulate 360 views of a similar resolution Catphan 504 phantom with a more powerful Nvidia RTX 2070 GPU. VOXSI is less computationally efficient than Fastcat, taking 45 seconds to simulate a CT with 512 $\times$ 512 $\times$ 1 voxels, 1024 detectors, and 780 projections, on two Intel Xeon 2.67 GHz processors. Simulations in DukeSim invoke a GPU MC tool to calculate scatter, the scatter calculation speed and total calculation speed thus depend on the desired noise. A sample DukeSim CT simulation with a phantom of 1900 $\times$ 1900 $\times$ 1000 voxels, 6912 projection images, and 47104 detectors took about 10 minutes on four Nvidia Titan Xp GPUs. In this case the simulation of such a large phantom and the added aspect of the MC noise is considered too different from Fastcat to compare computational efficiencies.

Fastcat introduces a new CBCT simulation method that could be a useful research tool. While MC simulations can agree with experimental results to the level shown in this work, they are computationally expensive. Fastcat is computationally efficient such that imaging settings can be quickly explored. Fastcat also contains a complex enough simulation method to agree closely with experimental results in terms of contrast and CNR, as seen in this work. Fastcat could be used as a quick and accurate exploration tool to examine the benefit of many CBCT imaging protocols. Further confirmation of promising protocols could then be performed using established MC and demonstrated with experimental methods.

Fastcat could also be used in a clinical setting: A Fastcat model of a given linac can be constructed such that it agrees with clinical CBCT quality assurance (QA) through the use of the virtual Catphan phantom. This can be done by adjusting the beam energy, focal spot size, detector material, and other parameters. Unlike with MC models, due to its fast simulation times Fastcat could build a scanner specific, virtual clinical datasets of hundreds of CBCT volumes of an anthropomorphic phantom can be constructed overnight. These datasets could be used to estimate the correct imaging settings for a given clinical situation. Such as estimating the CTDI necessary to achieve a given soft tissue contrast for a patient of a specific size. Likewise, Fastcat could be used in conventional workflows. Fastcat, with the addition of a mapping between experimental DICOM voxel values into material compositions like that of Schneider \textit{et al.} \cite{Schneider2000CorrelationDistributions}, could be used to simulate an MV CBCT from a kV CBCT, this would allow generation of MV beam's eye view images for the validation of a given treatment quickly from a given patient-specific CBCT acquired by the OBI.

Fastcat is unique in its simulation speed and accuracy. No other analytical CBCT simulator has shown such close agreement with experimental results, while no MC simulator has similar computational speed. Fastcat is limited, however, due to its dependence on pre-calculated data. Detector optical spread functions (OSFs), MV beams, and scatter kernels for different phantoms are stored in Fastcat. A simulation that uses a different detector, MV beam, or phantom than those available in Fastcat may need the additional step of simulating the necessary pre-calculated data in an MC application and be subject to new experimental validation. As an open-source simulator, ideally this new pre-calculated data could be uploaded by groups who use Fastcat, making a variety of commercial setups available and improving the simulator as a whole. Currently however, the type of accuracy shown in this work is only available for the Varian Truebeam EPID and Varian Truebeam OBI systems.

\section{Conclusion}

In this work we demonstrate experimental agreement between the Fastcat CBCT simulator and a Varian Truebeam linac. Fastcat 100 kVp and 6 MV CBCT images agreed closely with experimental kV and MV CBCT images, respectively. Close agreement was seen between experimental and Fastcat’s HU values, detector MTF, dose, and CNR.  CBCT simulation took 71 and 72 seconds for the kV and MV CBCT, respectively. We demonstrate the accuracy of a fast, free, open-source CBCT simulator that is well suited for the evaluation of new CBCT imaging protocols.

\section{Acknowledgments}

The authors would like to acknowledge Chelsea Dunning for the contribution to the name Fastcat. We would also like to thank Marios Myronakis for sharing the GOS MTF data with us. This research was enabled in part by support provided by WestGrid (www.westgrid.ca) and Compute Canada Calcul Canada (www.computecanada.ca). The work was partly funded by an NSERC Discovery Grant and the Canada Research Chairs program.

\section*{References}
\addcontentsline{toc}{section}{\numberline{}References}
\vspace*{-20mm}





\bibliography{./references}      



\bibliographystyle{./medphy.bst}    


\end{document}